\documentclass{andp2012}
\usepackage[english]{babel}

\usepackage{amsmath}
\usepackage{amsfonts}
\usepackage{amsthm}
\usepackage{graphicx}
\usepackage{dcolumn}
\usepackage{stackrel,amssymb}
\usepackage{color}
\usepackage{ulem}

\usepackage{bm}

\newcommand{\ket}[1]{\vert#1\rangle}
\newcommand{\bra}[1]{\langle #1\vert}

\usepackage{graphicx, psfrag, subfigure, color, amsmath, amsfonts, amssymb, appendix, setspace, rotating}
\newcommand{\be}{\begin{equation}}
\newcommand{\ee}{\end{equation}}

\newcommand{\Fig}[1]{Fig.~\ref{#1}}
\newcommand{\fig}[1]{fig.~\ref{#1}}
\newcommand{\Eq}[1]{Eq.~(\ref{#1})}

\newcommand{\bs}{\begin{split}}
\newcommand{\es}{\end{split}}

\usepackage{graphicx}
\usepackage{hyperref}

\usepackage{multirow,array}

\title{Quantum probing topological phase transitions by non-Markovianity}
\author[Gian Luca Giorgi]{Gian Luca Giorgi\inst{1,}\footnote{Corresponding author\quad E-mail:~\textsf{gianluca@ifisc.uib-csic.es}}}
\author[Stefano Longhi]{Stefano Longhi\inst{2,3}}
\author[Albert Cabot]{Albert Cabot\inst{1}}
\author[Roberta Zambrini]{Roberta Zambrini\inst{1}}
\address[1]{IFISC (UIB-CSIC), Instituto de Fisica Interdisciplinar y Sistemas Complejos Universitat de les Illes Balears-Consejo Superior de Investigaciones Cientificas, UIB Campus, E-07122 Palma de Mallorca, Spain}
\address[2]{Dipartimento di Fisica - Politecnico di Milano, Piazza Leonardo da Vinci 32, 20133 Milan, Italy}
\address[3]{Istituto di Fotonica e Nanotecnologie - Consiglio Nazionale delle Ricerche, Piazza Leonardo da Vinci 32, 20133 Milan, Italy}
\shortauthors{G. L. Giorgi et al.}
\begin{abstract}
Understanding the physical significance and probing the global invariants characterizing quantum topological phases in extended systems is a main
challenge in modern physics with major impact in different areas of science. Here, a quantum-information-inspired probing method is proposed where
topological phase transitions are revealed by a non-Markovianity quantifier. The idea is illustrated by considering the decoherence dynamics of an external
read-out qubit that probes a Su-Schrieffer-Heeger (SSH) chain with either pure dephasing or dissipative coupling. Qubit decoherence features and
non-Markovianity measure clearly signal the topological phase transition of the SSH chain.
 
\end{abstract}
\shortabstract
\begin{document}
\maketitle

\section{Introduction}

The discovery of the quantum Hall effect \cite{hall,haldane} 
 has revolutionized our understanding
of quantum phases of matter, with major impact in different areas of physics. Symmetry-protected topological phases are characterized by topological invariants and symmetries such that  
quantum phases with different topological invariants are not  connected to each  others by  perturbations, 
unless a spectral gap is closed. Remarkable examples are quantum Hall systems \cite{hall,haldane}, topological insulators, and topological superconductors \cite{bernevig,hasan,qi}.
 While in conventional Ginzburg-Landau theory phase transitions are identified by symmetry-breaking of a local order parameter \cite{landau}, topological phases are described by global invariants such as the Zak phase \cite{zak,berry},  the Chern number  \cite{nakahara,bott}, or the Thouless-Kohmoto-Nightingale-denNijs invariant \cite{tknn}.
 
 Since nonlocal topological invariants can escape a direct measure, identifying ways to relate topological invariants with measurable quantities has engaged physicists for long time.  The bulk-boundary correspondence \cite{qi}, relating the bulk topological invariants with the number of edge states in finite systems with open boundaries, 
provides the simplest route to measure topological numbers \cite{Goldman,Mittal,Feng}. However, edge states are not always accessible and the bulk-boundary correspondence can even fail in certain models, such as in non-Hermitian systems \cite{kunst18,Yao18}. Direct measurements of topological invariants in the bulk have been proposed and demonstrated in a series of experiments with synthetic matter, such as those based on Bloch oscillations \cite{atala,ramasesh}, unitary and non-unitary quantum walks \cite{kitagawa1,kitagawa2,zeuner,cardano,lewenstein,zhan17,Meier18},  and out-of-equilibrium (quench) dynamics \cite{caio,hu,wilson,song,tarnowski,wang,zhang,sun}. 
In one dimension, the simplest model exhibiting a topological phase transition is the celebrated Su-Schrieffer-Heeger (SSH) model, describing a tight binding 
1D lattice with staggered hopping amplitudes \cite{ssh,TIbook}. The SSH model exhibits two distinct topological phases characterized by different values of Zak phase \cite{atala}. Such topological phases have been experimentally observed in different physical settings, including cold-atom platforms \cite{atala},   
 photonic systems \cite{kitagawa2,zeuner,lewenstein} and topological circuits \cite{Liu19,Liu19b}. Recently, a great attention has been devoted to study topological phases in open quantum and classical systems (see, e.g., refs. \cite{RRef3a,RRef3b,RRef3c,RRef3d,RRef3e,RRef3f,RRef3g,RRef3h} and references therein), and several unusual effects have been disclosed such as strong sensitivity to boundary conditions and breakdown of the Bloch bulk-boundary correspondence owing to the non-Hermitian skin effect \cite{RRef3i,RRef3l,RRef3m,RRef3n,RRef3o,RRef3p}.
 
 \par
The identification of topological phases  by means of quantum-information-oriented indicators is a rather new and exciting area of research, which remains largely unexplored.  
Previous studies highlighted the interplay between topology and entanglement entropy, entanglement spectrum and multipartite entanglement \cite{
Kitaev06,Levin06,Li08,pezze}. In particular, entanglement was used as a probe in the spin-$1/2$ Heisenberg model on the kagome lattice   \cite{jiang} and in the Kitaev model \cite{nori} while, for the SSH model, bipartite ground-state  entanglement was shown to exhibit a 
nonanalytical behavior around the topological phase transition \cite{cho}. In the context of open quantum systems and many-body physics, quantum coherence of probes and Loschmidt echo are very useful quantitates to detect phase transitions is different settings. Among others, we mention Loschmidt echo enhancement by  quantum criticality and dynamical quantum phase transitions in the transverse-field Ising model \cite{RRef3q,RRef3r}, dynamical topological quantum phase transitions in the postquench time evolution of quantum many-body systems initialized in mixed states \cite{RRef3s}, and the detection of critical times and Lee-Yang zeros in probe spin coherence dynamics \cite{RRef3t,RRef3u,RRef3v}.
\\
One of the most important topics recently developed in the field of open quantum systems is certainly the search for memory effects in large environments.  Then, a variety of approaches has been proposed to quantify the amount of non-Markovianity focusing on Markovian approximations in open quantum systems, the characterization of memory effects and  their quantitative estimate by non-Markovianity measures 
\cite{NM_Rivas,NM_Vacchini,NM_Vega,NM_hierarc,NM_chen,NM_zhang,NM_strasberg}. This is because a precise knowledge of the environmental properties can be exploited to mitigate dissipation or to exploit it by engineering the bath \cite{cirac,nokkala}.

In this paper, we propose a quantum probing scheme based on an external read-out 
qubit where  topological phase transitions are revealed by non-Markovianity measure. While previous bulk probing methods require rather generally to access the dynamical evolution in the lattice \cite{atala,kitagawa2,zeuner,lewenstein}, in our setting the probing is fully external and topological phases are retrieved by the decoherence dynamics of the qubit. 
In open quantum systems, non-Markovian quantum probes have been recently 
explored to extract information from  
environments \cite{Vasile,nokkala,Haikka,NM_Mott,NM_And,NM_And2}. In particular, local and global 
external probes of  many-body systems have been considered  looking at their
non-Markovianity in connection with
quantum phase transition in Ising models \cite{Haikka}, 
superfluid-to-Mott-insulator transition \cite{NM_Mott} and Anderson localization in 
disordered environments \cite{NM_And}. 

 Our main idea is illustrated by considering decoherence dynamics of a qubit locally probing an SSH lattice by either pure dephasing or dissipative interaction. 
A sharp transition between non-Markovian and 
Markovian dynamics is observed when the SSH gap closes and reopens into a different topological phase. 
The local interaction between the chain and the probe gives rise to new 
states, localized around the probe [bound states (BSs)], the number and the 
symmetry properties of which are 
strongly dependent on the  topological phase of the SSH chain. This enables the 
sharp transition between non-Markovian and Markovian dynamics of the 
(out-of-equilibrium) probe. For a dephasing qubit the relation between
Loschmidt echo and the information flow between the probe and the system was reported in ref.\cite{Haikka}.
This read-out of the topological phase is robust, as different measures of non-Markovianity coincide for dephasing probe 
and, furthermore, it does not need fine tuning of probing strengths.  
On the other hand the dissipative qubit scheme allows to read-out the topological phase  with an even sharper transition but 
its operation is limited to a restricted probing strengths regime.
The paper is organized as follows. Section \ref{sec2} provides the description of the SSH model and of its dephasing interaction with the external qubit used as a probe. The decoherence dynamics of the probe and the exact energy spectrum of the coupled qubit-bath system are presented in Section \ref{sec3}. The non-markovian features of the probe dynamics are  discussed in Section \ref{sec4}, where we show how the topological phase transition of the SSH chain can be revealed by  a non-Markovian quantifier. The main conclusions are outlined in Section \ref{sec5}. Finally,  some technical details and a different coupling scheme are presented in two Appendices.

\section{Model}\label{sec2}

\begin{figure}
\begin{center}
\includegraphics[scale=.43]{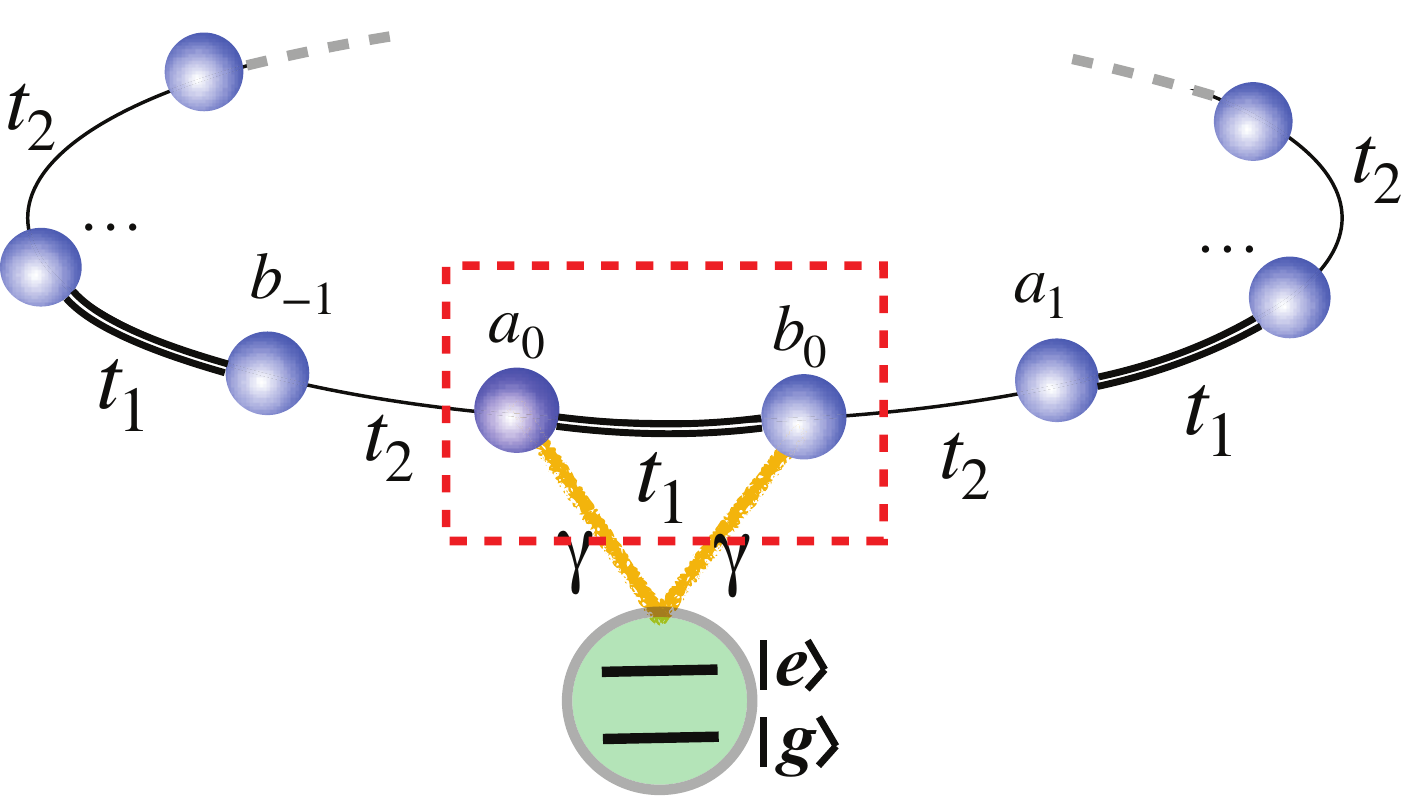}
\end{center}
\caption{ (Color online) Schematic view of the probing model. A cell of the SSH chain is coupled to an external qubit whose dephasing dynamics is monitored to detect topological phases of the chain.   }\label{figssh}
\end{figure}

 We consider a 1D SSH lattice probed by a qubit interacting with the unitary cell of the lattice  (Figure~\ref{figssh}). Indeed, if the interaction were limited to a single site, symmetry reasons would rule out any difference between the two topological phases.   The Hamiltonian of the full system is given by
\begin{equation}\label{eq1}
H=H_{SSH}+H_I+H_P
\end{equation}
 where $H_P$ is the Hamitonian of the  probe qubit,
\begin{equation}\label{eq2}
H_{SSH}=t_1\sum_{n=1}^N(a_n^\dag b_n+hc)  +t_2 \sum_{n=1}^N(a_n^\dag b_{n-1}+hc),
\end{equation}
is the Hamiltonian of the SSH chain,  with $a,b$ either fermionic or bosonic annihilation operators  and $H_I$ describes the qubit-chain interaction. The SSH chain comprises $N$ unit cells, and periodic boundary conditions are assumed. The SSH chain is in the topologically nontrivial phase for $t_2>t_1$, while it is in the trivial topological phase for $t_2<t_1$: indeed, under open boundary conditions only in the former case the chain sustains topologically-protected edge states.
As for interaction term, it can be either of dissipative or dephasing nature  for the probe. 
Here we will primarily focus on the dephasing scenario, as it allows for a direct link between topology detection and non-Markovianity. Signatures of the topological phases can be also found looking at the non-Markovianity of the probe dynamics in the dissipative case, but the results there are less direct and will be described in detail in Appendix \ref{appA}.

The interaction Hamiltonian in the  pure dephasing case, (where the probe Hamiltonian can be written as $H_P=(\omega_0/2)\sigma_z$, with  $\sigma_z=\ket{e}\bra{e}-\ket{g}\bra{g}$  the pseudospin operator associated to the probe qubit), analogous to what normally used in quantum nondemolition protocols,  can be obtained working in the dispersive coupling limit:
 if we denote with and we label  with $0$ the chain cell at which the probe is coupled, 
\begin{equation}\label{eq3}
H_I=\gamma\frac{\mathbb{I}+\sigma_z}{2} \otimes (a_0^\dag a_{0}+b_{0}^\dag b_{0})
\end{equation}
 where $\gamma$ measures the coupling strength. This $\sigma_z\otimes n$ interaction has been reported in the $1990$s for atom-light interaction in the dispersive
limit \cite{brune,brune2}. In these experiments, Rydberg atoms interact with microwave photons far from resonance, in
high-Q superconducting cavities, and this coupling models the energy shift proportional to the number of signal photons. Dispersive couplings have also been reported in dispersive optomechanics for membranes in cavities when placed at the nodes of the light field \cite{thompson,Jayich} or in cavity quantum
electrodynamics for superconducting electrical circuits \cite{blais}.

\section{Decoherence dynamics} \label{sec3}

Taking as initial state at $t=0$ a coherent state  for the probe $C_e \ket{e}+C_g\ket{g}$ and a generic chain state $\ket{\Psi_0}$ (that will be chosen in the one-excitation sector), the evolution of the probe reduced density matrix reads
\begin{equation}
\rho_p(t)=\left(
\begin{array}{cc}
|C_e|^2& C_e C_g^* q(t)\\
C^*_e C_g q^*(t)&|C_g|^2
\end{array}
\right)
\end{equation}
where the only time-dependent coefficient is given by
\begin{equation}\label{echo}
q(t)=\bra{\Psi_0}e^{iH_{SSH}t}e^{-i \bar H t}\ket{\Psi_0},
\end{equation}
with $\bar H=H_{SSH}+ \gamma (a_{0}^\dag a_{0}+b_{0}^\dag b_{0})$.  Its modulus square  $L(t)=|q(t)|^2$ is known as Loschmidt echo (LE) and is a direct measure of decoherence. This is typical of pure dephasing dynamics, where the  probe decoherence rate is fully determined by a quantity that only depends on environmental degrees of freedom.
In order to  determine the dynamics of the LE, it is then necessary to study the properties of the renormalized Hamiltonian $\bar H$, which corresponds to an effective SSH chain displaying a potential-site defect at the $n=0$ unit cell. For a sufficiently large number of sites,  $\bar H$ keeps the two-band structure of $H_{SSH}$. However, up to four BSs around the impurity can emerge that can deeply affect the probe dynamics. A way to determine the presence and the shape of BSs is based on a simple Ansatz that exploits the mirror symmetry of $\bar H$. If such states exist, their wavefunction \textit{must}  be either even or odd around the impurity cell $0$, that is, they can be written  as
  \begin{equation}\label{eqb}
 \ket{\psi}_\pm = [c_0 (a_{0}^\dag\pm b_{0}^\dag ) +c_1  (b_{-1}^\dag \pm a_{1}^\dag )+ c_2  (a_{-1}^\dag \pm b_{1}^\dag )+\dots]\ket{0}
 \end{equation}
From the time-independent Schr\"odinger equation,  these two families of states lead to two sets of coupled
equations that can be solved analytically in the thermodynamic limit through a scaling Ansatz. Taking
into account the  structure of the model, we can introduce the dimeric localization parameter $X$ and the dimer imbalance parameter $Y$ such that
 \begin{eqnarray}
 c_{2n}=c_0 X^{-n} & \qquad n \geq 1 \\
c_{2n+1}=c_{2n} Y  & \qquad n \geq 0
 \end{eqnarray} 
 with $|X|>1$ for localization. The eigenenergy and localization/imbalance parameters are found as solutions of coupled algebraic equations; details are given in Appendix  \ref{appB}. 
Depending on the ratio $t_1/t_2$, we face two different scenarios:  (i) For $t_1<t_2$ (topological non-trivial phase), there are always two symmetric BSs, irrespective of the strength of the coupling constant $\gamma$. As for the anti-symmetric equations, we have two solutions for $t_1<\gamma/2$, while no solutions are found whenever $t_1>\gamma/2$. (ii) For $t_1>t_2$ (topological trivial phase), there exist one symmetric and one anti-symmetric solutions, without any dependence on $\gamma$.\\
The band structure of $\bar H$ and diagram of BSs is presented in Figs. \ref{figbands}(a) and (b) for the two topological phases as a function of the coupling $\gamma${; the behavior of the localization parameter $X$ as a function of $t_1/t_2$, which determines the boundary of existence of BSs, is given in Appendix \ref{appB}.
 The energy spectrum is composed by the  two minibands  of the SSH lattice, with gap separation $|t_1-t_2|$, and a number of  BSs (point spectrum) with energies either in the gap or above the upper miniband (\Fig{figbands}). As commented above, the  symmetry properties of such BSs change drastically across the critical point $t_1/t_2=1$: irrespective of the value of $\gamma$, in the phase
 $t_1<t_2$, there is an even number of symmetric BSs and an even number of antisymmetric BSs, while both 
 symmetric and antisymmetric BSs appear in an odd number in the phase $t_1>t_2$.

In the  case $\gamma< 2 t_1$, which is the most typical operational regime, the qubit-SSH system sustains two BSs. The two BSs are both symmetric in  the topological phase $t_2>t_1$, whereas in the topological phase $t_2<t_1$ they have opposite parity. This is because, at the gap closing point $t_2=t_1$, the parity of the in-gap  bound state changes from symmetric to anti-symmetric [see \Fig{figbands}(c)]. The flip of parity of the in-gap bound state at the band closing point is a topological property, related to the topological phase transition of the SSH chain, and it turns out to be robust against off-diagonal disorder in the chain, as shown in the next section. Indeed, the two bound states present for $\gamma< 2 t_1$ (assuming $\gamma>0$, which makes the defect Hamiltonian nonnegative) directly derive from  the two upper band-edge states of the SSH chain, which are expelled from their respective bands in the presence of the renormalization induced by the probe, as their energies are lifted up. These band-edge states have energies respectively $|t_1+t_2|$ and $-|t_1-t_2|$, as it can be obtained from equations (\ref{ep}) and (\ref{em}) taking $\gamma\to 0$. As the cell defect does not break the spatial mirror symmetry of the model,  the new BSs \textit{must} belong to the same symmetry sector of their zero-defect limit states. Now, while the band-edge state with energy $|t_1+t_2|$   is completely uniform along the chain and symmetric independently on the topological phase of the SSH model, the state with energy $-|t_1-t_2|$ is a symmetric one for $t_1>t_2$ while is antisymmetric for $t_1<t_2$. This change of symmetry directly originates in the SSH winding number, as it is critically related to the eigenvalues in the momentum space. Thus, the flip of parity of the in-gap bound state is due to the change of topological phase  of the SSH chain. In the context of different topological models, a dependence of defect-induced in-band BSs on the symmetries of the model was also  reported in refs. \cite{prl_def,prb_def}.

\begin{figure}
\includegraphics[scale=.3]{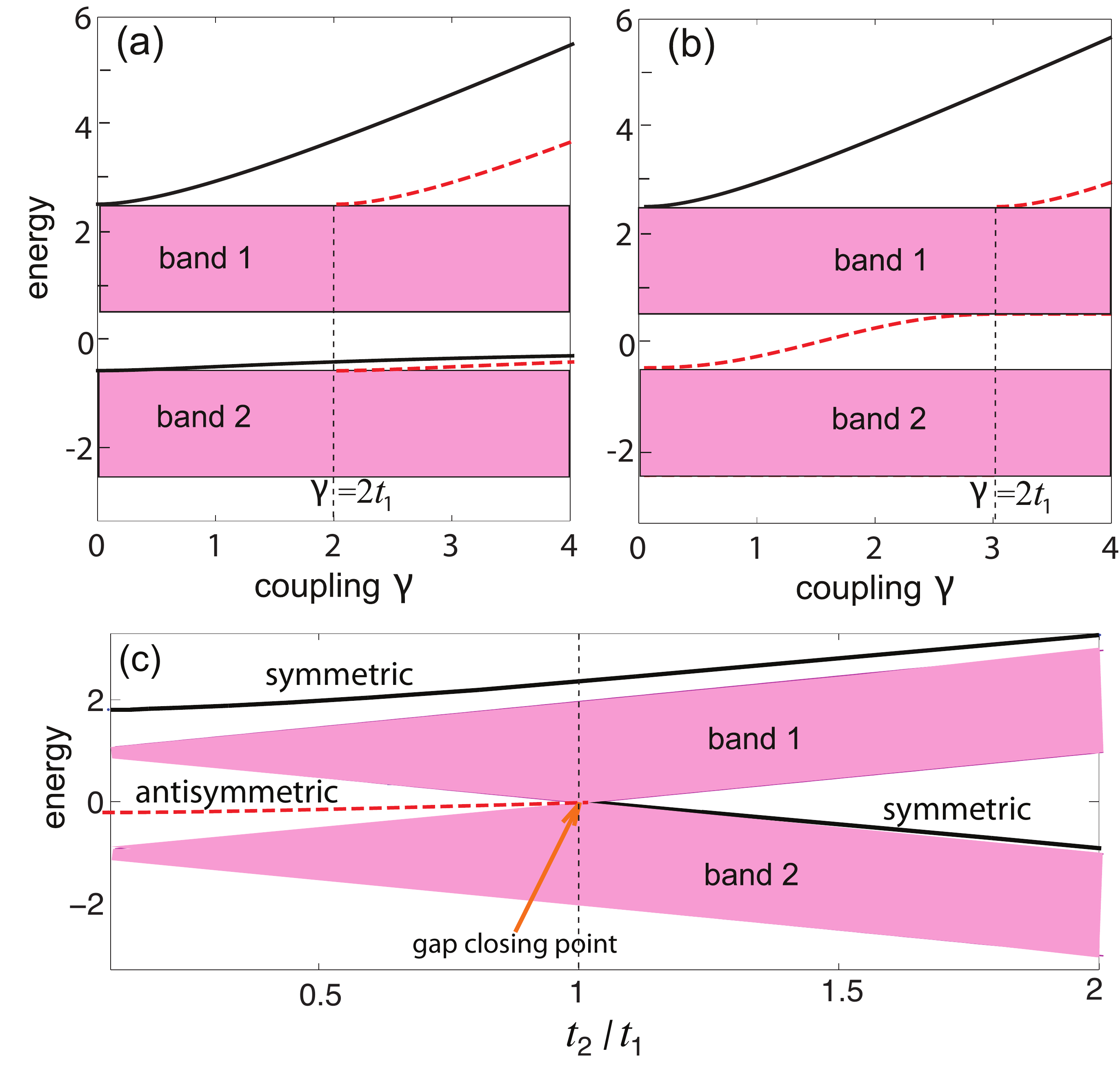}
\caption{  (Color online)
Energy spectrum of $\bar{H}$ as a function of the coupling $\gamma$ for $t_1=1 $ and $t_2=1.5$ [panel (a)] and for $t_1=1.5 $ and $t_2=1$ [panel (b)]. The pink (gray) bands represent the continuum spectrum and are those of the bare $H_{SSH}$ chain. BSs are denoted by solid lines if they are symmetric  or by dashed lines if they are anti-symmetric (see main text). (c) Energy spectrum of $\bar{H}$ as a function of the coupling ratio $t_2 / t_1$ in the regime $\gamma< 2 t_1$ ($\gamma=0.8 t_1$ in the plot). At the gap closing point $t_2=t_1$, the parity of the in-gap bound state flips from symmetric to anti-symmetric. 
 }\label{figbands}
\end{figure}

\section{Non-Markovianity Measure and Topological Phase Transition} \label{sec4}

The presence or the absence of BSs and their symmetry properties deeply change the decoherence dynamics of the qubit. In particular, the parity flip of the in-gap BS at the band closing point can be exploited to reveal the topological phase transition of the SSH bath. In fact, depending on the system parameters and on the initial state of the chain, the coherence (\ref{echo}) can experience either a pure exponential decay or oscillatory behavior. In particular, we are interested in studying the behavior of the coherence and LE around the topological critical point $t_1=t_2$  and to relate the phase transition to the non-Markovianity measure. Let us assume that the initial state of the lattice is an anti-symmetry state $\ket{\Psi_0^A}$ around the impurity cell. The choice of such a state is motivated by the fact that its symmetry properties makes it especially sensitive to the topological phase, however we stress that our results hold even though the initial state is not perfectly antysimmetric, i.e. our probing protocol is robust against unavoidable imperfections in the initial state preparation of the system.  The dynamics of $\ket{\Psi_0^A(t)}$ will be greatly affected by the presence of BSs and by their symmetries, which are in turn established by the topological phase of the SSH bath. In fact,  for $t_1<t_2$,  $\ket{\Psi_0^A}$ is orthogonal to both BSs, that are indeed symmetric, while for $t_1>t_2$  one of the two BSs will contribute to the dynamics  (see \Fig{figbands}), which will eventually give rise to an oscillatory behavior of $L(t)$. Since the symmetry flipping of the in-gap BS requires the gap to close and reopen in the other topological phase of the SSH chain [\fig{figbands}(c)], the qualitative change of the LE, from exponential-like to oscillatory-like, is thus of topological origin. Examples of the decay dynamics of the coherence $|q(t)|$ in the two different topological phases are given in the inset of \Fig{fignm}(a). In the simulations, the initial state belongs  to the one-excitation sector
\begin{equation}
 \ket{\Psi_0^A}=(1/2)(a_{0}^\dag-b^\dag_{0}+a_{1}^\dag-b_{-1}^\dag )\ket{0},
\end{equation}
 corresponding to an anti-symmetric state homogeneously distributed over two unit cells, and the time evolution is computed in the two topological phases for a fixed value of the coupling $\gamma/t_1$. For $t_2>t_1$ (topological non-trivial phase), the dynamics of $\ket{\Psi_0^A}$ is blind with respect to the two BSs and only depends on the interaction with the continuous bands. Then, after a very small initial bounce, the system will end up in an equilibrium steady state. On the other hand, for $t_2<t_1$ topological trivial phase),the anti-symmetric BS of $\bar H$ will significantly influence the dynamics resulting in the macroscopic oscillation of the coherence. It is important here to remark that this behavior is independent of a fine tuning of $\gamma$ (which needs  to satisfy $\gamma<2 t_1$), that would only affect the details of  $q(t)$.  In other words, a sharp transition between  stationary and oscillating regimes is always observed, regardless of the coupling strength $\gamma$, since it is fully determined by topology.

\begin{figure}
\includegraphics[scale=0.75]{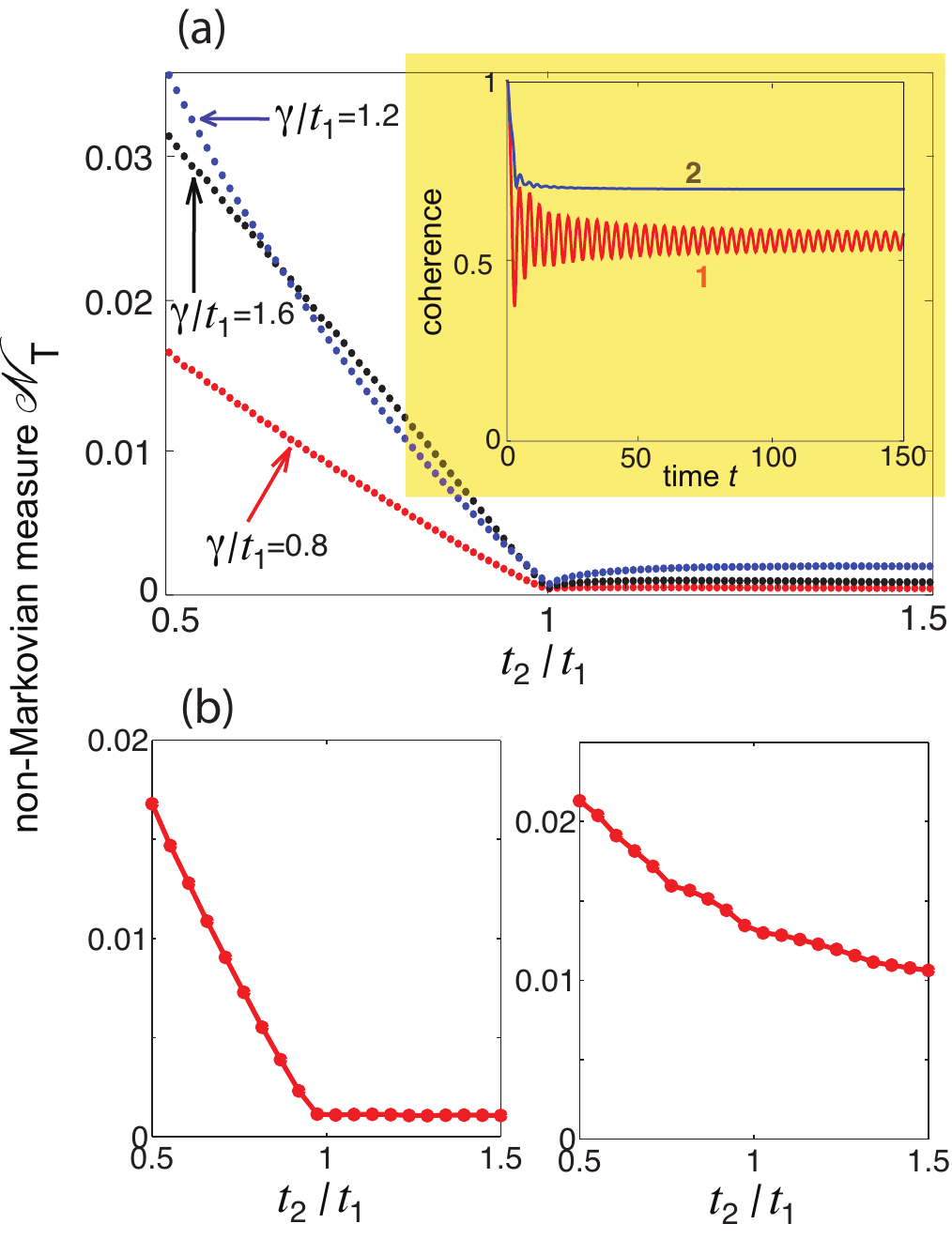}
\caption{(Color online) (a) Numerically-computed non-Markovianity quantifier ${\cal N}_T$ ($T=150\; t_1$) as a function of $t_2/t_1$. The initial state is $\ket{\Psi_0^A}=(1/2)(a_{0}^\dag-b^\dag_{0}+a_{1}^\dag-b_{-1}^\dag )\ket{0}$. The behavior of  ${\cal N}_T$ is shown for three different values of the ratio $\gamma/t_1$ [$\gamma/t_1=0.8$ (red), $\gamma/t_1=1.2$ (blue), and $\gamma/t_1=1.6$ (black)]. In all cases a clear phase transition is observed at $t_1=t_2$, with ${\cal N}_T$ almost vanishing in the $t_2>t_1$ topological phase. Longer times would slightly modify the value of ${\cal N}_T$, as the initial oscillations would weight less. The inset shows the detailed dynamics of the coherence $|q(t)|$ in the two distinct topological phases for the same coupling strength  $\gamma=1.2 t_1$. Curve 1: $t_2=0.6 t_1$ (red); curve 2:  $t_2=1.4 t_1$ (blue). A chain of $N=200$ cells has been considered in the numerical simulations. (b) Numerically-computed non-Markovianity quantifier ${\cal N}_T$ versus $t_2/t_1$ for $\gamma / t_1=0.8$ and for two different initial excitation states $\ket{\Psi_0}$. In the left panel $\ket{\Psi_0}=(1/ \sqrt{3})(\hat{a}^{\dag}_0-\hat{b}^{\dag}_0+\hat{a}^{\dag}_1)|0 \rangle$, which does not have any special symmetry, whereas in the right panel $\ket{\Psi_0}=(1/ \sqrt{2}) (\hat{a}^{\dag}_0+\hat{b}^{\dag}_0) | 0 \rangle$ is a symmetric state.}
 \label{fignm}
\end{figure}

 The  oscillations of the LE correspond to a periodic reflux of information from the bath back to the probe, which can be studied in terms of non-Markovianity through, for instance, the quantifier $\mathcal{N}$ of ref.\cite{nm},  defined as
 \begin{equation}
 \label{eq:N}
\mathcal{N}=\max_{\rho_1,\rho_2}\int_{\sigma>0}dt\;\sigma(t,\rho_{1,2}(0)),
\end{equation}
where 
$\sigma[t,\rho_{1,2}(0)]=dD[\rho_{1}(t),\rho_{2}(t)]/dt$  is the rate of change of the trace distance $D(\rho_{1},\rho_{2})={\rm Tr}|\rho_{1}-\rho_{2}|/2$, which quantifies distinguishability. The maximum in Equation
\eqref{eq:N} is taken over any possible pair of initial states $\{\rho_1,\rho_2\}$. 
Such a measure of non-Markovianity  stems from the observation that, under a dynamical semi-group (and then under a quantum Markov process), the trace distance of any pair of initial states is a monotonically decreasing function of time, which means that the distinguishability between pairs of quantum
states can only decrease (contractiveness). 
This loss of distinguishability is due to a  flow of information from the
system to the environment.
If such property is violated, then the dynamics cannot be Markovian. 
Thus, the existence of time
windows where this contractive property is violated
witnesses the presence of non-Markovianity in the dynamical map.
In the case of a purely decohering dynamics of a qubit, the only time-dependent parameter entering the evolution is $L$. The violation of contractiveness is witnessed by the increase of such quantity and then of the qubit coherence \cite{he}. 
The amount of such violation  can be directly calculated 
as the integral of $d \sqrt{L(t)}/dt$ extended over the time intervals where the derivative $\dot L$ is positive.
  It is particularly convenient that, actually, in the presence of pure dephasing, different measures of non-Markovianity coincide \cite{nm,NM_Rivas,NM_hierarc}, as they only depend on a single time-dependent function (the LE in our case). This also implies that $\mathcal{N}$ is easily accessible in experiments, as shown for instance in ref.\cite{piilo} in the case of photons.

   As already pointed out in  refs. \cite{francesco,NM_And}, ${\mathcal N}$   diverges if $|q(t)|$ exhibits  oscillations, as it  happens in our case.  This drawback can be simply circumvented by time average   ${\mathcal N}$  over several periods of oscillations: we can in fact define 
\begin{equation}
{\mathcal  N}_T= \frac{1}{T} \int_{0 \; \dot{L}>0}^{T} dt \frac{d \sqrt{L(t)}}{dt}.
\end{equation}
 From an experimental point of view, using ${\mathcal  N}_T$  instead of ${\mathcal  N}$ has no physical implications, as the probe dynamics has to be measured in finite time intervals in both cases.  In  \Fig{fignm}(a),  ${\mathcal  N}_T$ is displayed for three different chain-probe coupling constants as a function of $t_2/t_1$. The topological transition observed is continuous around the critical point, as it is determined by the spectral weight of the antisymmetric BS with respect to the initial state. This quantity is exactly zero for $t_1=t_2$ and then starts  building up in a continuous fashion as the BS gets more localized.   The discontinuity in the  derivative (as in ordinary first-order phase transitions) of ${\mathcal  N}_T$ showed in \Fig{fignm}(a) tells us that non-Markovianity provides a sharp indicator that can be used as an effective order parameter to identify the topological phase transition. Remarkably, the discontinuous behavior of the derivative of ${\mathcal  N}_T$,  as the  topological SSH phase is crossed, is observed for rather  arbitrary initial preparation of the lattice, that could deviate from an exact anti-symmetric state [see Figure3(b)]. Only for an exact symmetric initial state the non-Markovian measure is not able to predict the topological phase transition [right panel in Figure3(b)].

The topological protection of edge states in the SSH model against off-diagonal disorder, i.e. disorder in the hopping amplitudes that respects the chiral symmetry of the system, is a well established feature. Likewise, in our qubit-bath system we argue  that the symmetry flip of the in-gap BS at the gap closing point, shown in \Fig{figbands}(c), is not destroyed by moderate disorder in the system that does not break chiral symmetry. This means that our probing protocol, based on the non-Markovian measure $\mathcal{N}_T$, is robust against imperfection in the system.
In order to test the robustness of our probing protocol, we numerically-computed   ${\mathcal  N}_T$ versus $t_2/t_1$ when the qubit is coupled to a disordered SSH chain, i.e. with inhomogeneous values of the ratio $t_2/t_1$. In the numerical analysis, disorder has been simulated by assuming $t_1=\langle t_1 \rangle (1+\delta t_1)$ for the intra-dimer hopping amplitude, where $\langle t_1 \rangle$ is the mean value of $t_1$ and $\delta t_1$ a random variable with uniform distribution in the range $(-\delta,\delta)$. Clearly, in the disordered system the parity (odd/even) symmetry of the eigenstates of $\bar{H}$ is lost because spatial inversion symmetry around the qubit site is broken, so that BSs can not be classified anymore as symmetric or antisymmetric states. Nevertheless, we can introduce an indicator of the symmetry of an eigenstate about the defect unit cell via a parity index $\mathcal{P}$ defined as follows 
\begin{equation}
\mathcal{P}=\frac{1}{2} \frac{ \sum_n \left| a_n+b_{-n}\right|^2 }{\sum_n \left(  |a_n|^2+|b_n|^2 \right)}
\end{equation}
with $0 \leq \mathcal{P} \leq 1$, $\mathcal{P}=1$ for an exact symmetric state ($a_n=b_{-n}$) and $\mathcal{P}=0$ for an exact anti-symmetric state ($a_n=-b_{-n}$). Hence an eigenstate with parity index $\mathcal{P}$ close to one means that the eigenstate is dominantly a symmetric state, whereas an eigenstate with $\mathcal{P}$ close to zero indicates that it is a dominant anti-symmetric function. Figure 4 shows the numerically-computed behavior of the energy spectra and non-Markovianity quantifier $\mathcal{N}_T$ versus $t_2/ \langle t_1 \rangle$ in a disordered SSH chain, comprising 301 unit cells, for two different values of disorder strength and for $\gamma=0.8 \langle t_1 \rangle$. The initial state is $\ket{\Psi_0^A}=(1/2)(a_{0}^\dag-b^\dag_{0}+a_{1}^\dag-b_{-1}^\dag )\ket{0}$. An inspection of the energy spectra indicates that, even for a moderate disorder in chain, the BSs induced by coupling to the qubit keep a high degree of parity, as measured by the parity index $\mathcal{P}$ [see the insets in Figs.4(b) and (e)], and that the parity of the in-gap BS is flipped as the gap closes and reopens at $t_2= \langle t_1 \rangle$.
The main difference as compared to the ordered lattice is that the value of ${\mathcal  N}_T$  in the $\langle t _1 \rangle<t_2$ topological phase does not vanish but settles down to a steady value that increases as the disorder strength increases. This behavior is likely to be ascribed to Anderson localization of bulk states in the lattice, which enhances non-Markovianity \cite{NM_And}. Such an enhancement  is featureless as $t_2/ \langle t_1 \rangle$ is varied, so that the signature of the topological phase transition is still clearly visible and the effect of disorder is basically to provide a bias to $\mathcal{N}_T$, as shown in Figs.4(c) and (f).
\begin{figure*}
\begin{center}
\includegraphics[scale=.7]{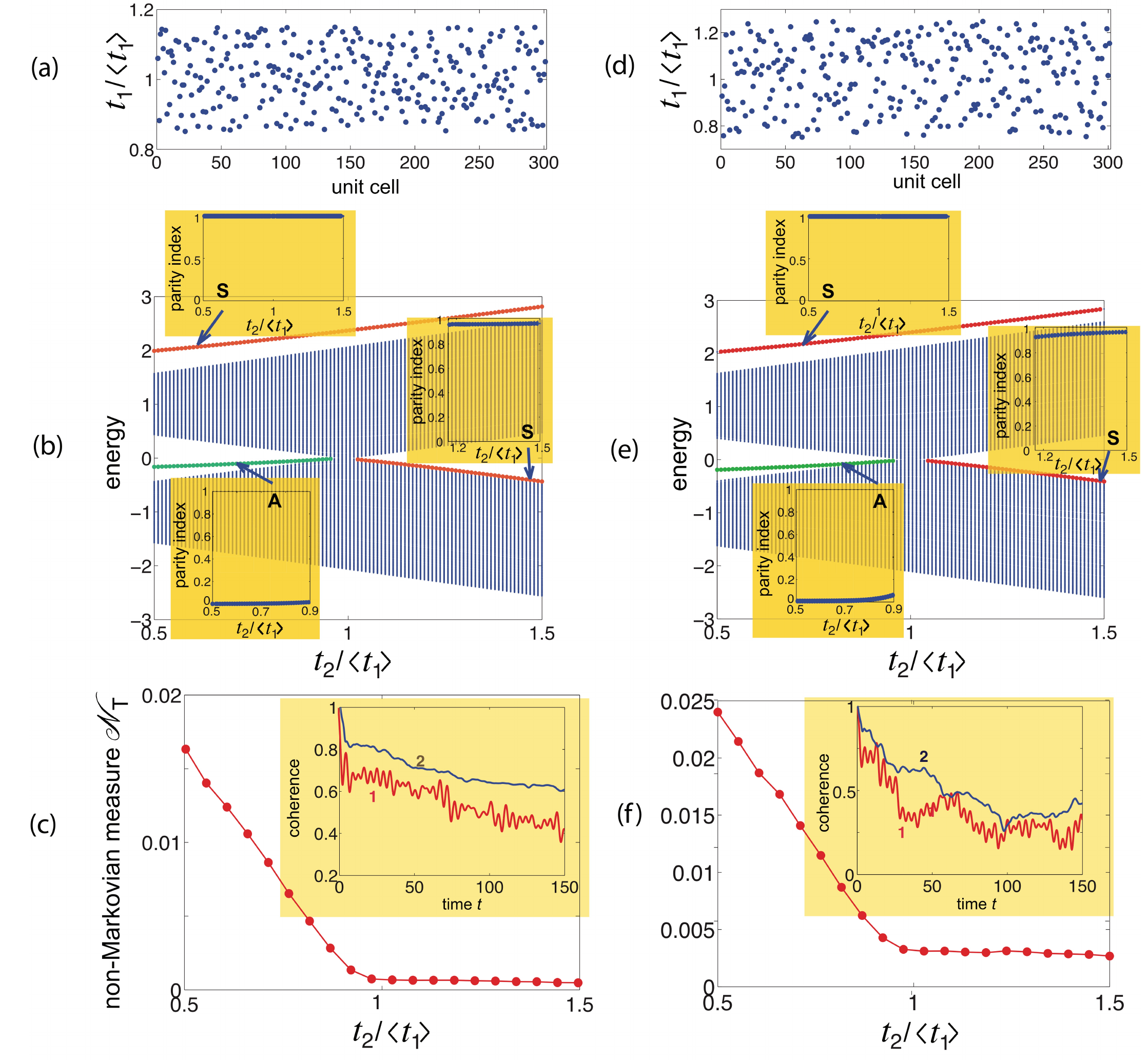}
\caption{(Color online) Effect of hopping disorder on energy spectra and non-Markovianity quantifier $\mathcal{N}_T$. (a) Behavior of hopping rate $t_1$ versus number of unit cell in the lattice for a realization of disorder (disorder strength $\delta=0.15$). The qubit-bath coupling is $\gamma=0.8 \langle t_1 \rangle$. (b) Numerically-computed energy spectrum of $\bar{H}$ versus $t_2 / \langle t_1 \rangle$. The insets show the behavior of the parity index $\mathcal{P}$ of the two BSs, indicating that parity flipping at the gap closing point is still observed in the presence of disorder. (c) Numerically-computed non-Markovianity quantifier ${\cal N}_T$ ($T=150\; \langle t_1 \rangle$) as a function of $t_2 / \langle t_1 \rangle$. The inset shows the detailed dynamics of the coherence $|q(t)|$ in the two distinct topological phases for $t_2=0.6 \langle t_1 \rangle$ (curve 1, red) and  $t_2=1.4 \langle t_1 \rangle $ (curve 2, blue). (d-f) Same as (a-c), but for a disorder strength $\delta=0.25$. }
\end{center}
\end{figure*}

Before concluding, let us remark that  the connection between non-Markovianity and probing of topological phases is strengthened by the fact that it is not strictly related to the model discussed so far but can be also found in the presence of a very dissimilar probe-chain interaction (dissipative coupling).  In fact, as detailed in Appendix \ref{appA}, also in this different probing protocol, characterized by the interaction Hamiltonian $ H_I= \gamma[\sigma_+ (a_0+b_0)+ \sigma_{-} (a^{\dag}_0+b^{\dag}_0)]$ \cite{tudela},  it is possible to detect the topological phase transition by looking at the behavior of ${\mathcal  N}_T$.
The main difference with respect to the dephasing case is that an abrupt transition in the non-Markovianity measure only takes place if we set $\gamma=t_1$. Otherwise, ${\mathcal  N}_T$
cannot be taken as an order parameter and  a direct inspection of the  shape of the output would be necessary to infer the topological phase (see Appendix \ref{appA}). Indeed, the number of BSs would in any case determine the number of frequencies observed in the output signal, which could be detected by a spectral analysis.
 This feature  can also be found in the dephasing scenario in the case where initial state is not an antisymmetric one. Indeed, the dependence of the parity of the number of BSs on the topological phase of the SSH chain would in any case warrant qualitatively different probe dynamics and then different spectral contents.

\section{Conclusions }\label{sec5}
  Non-Markovianity and revivals of quantum coherence have attracted a great attention in the past recent years as potential resources in different contexts \cite{NM_Rivas,NM_Vacchini,NM_Vega,NM_hierarc}.  Here, we have shown how non-Markovianity can be harnessed to externally probe the topological phases of quantum matter, without the need to detect the dynamics of the entire system.  We have proposed a quantum-information-inspired strategy to measure global topological invariants of extended quantum systems  by monitoring the decoherence dynamics of  an external probe  locally coupled to the system. The main idea  has been illustrated by considering the non-Markovian dynamics of an external read-out qubit probing an SSH chain both for dephasing and dissipative reduced dynamics. 
Exploiting the bound-state phase diagram, one can choose a convenient set of possible initial states and tailor a very efficient probing protocol. Indeed, 
it turns out that non-Markovianity does not vanish solely in one of the two topological phases of the chain, thus revealing its topological order and serving as a sharp indicator for the transition between different phases. 
While we have studied the single-excitation regime for the SSH chain, it would be interesting to analyze the problem in the general case of a Fermi or a Bose gas, where the symmetry diagram of the BSs, which  stems from the renormalized Hamiltonian, could still play a relevant role.

 The system we have described is suitable for experimental implementation both in photonic and cold-atom platform, where the SSH chain can be simulated, as well as its coupling to the external probing qubit \cite{tudela}. As commented above, the dephasing dynamics can be achieved working in the dispersive limit (strong detuning). 
Our results, besides of shedding new light onto the topological significance of memory effects in the dynamics of open quantum systems, suggest that topological phases can provide a powerful means to control (either enhance or suppress) quantum decoherence.

\begin{acknowledgements}
The authors acknowledge support from  MINECO/AEI/FEDER through projects EPheQuCS FIS2016-78010-P and the Mar\'ia de Maeztu 
Program for Units of Excellence in R\&D (MDM-2017-0711),  funding from ``Conselleria d'Innovaci\'o, Recerca i Turisme del Govern de les Illes Balears" PhD and postdoctoral programs, and 
the UIB ``professors convidats" program. 
\end{acknowledgements}

\appendix

\begin{center}
\section*{ Appendix }
\end{center}
\renewcommand{\theequation}{A-\arabic{equation}}
\setcounter{equation}{0}
\renewcommand\thefigure{A-\arabic{figure}} 
\setcounter{figure}{0}

\section{Dissipative Model}\label{appA}
We consider a two-level system (qubit) as a probe of the topological phase, dissipatively coupled to the SSH lattice  as schematically shown in \Fig{figssh}.
The model of the qubit dissipatively-coupled to an SSH lattice is defined by the Hamiltonian \Eq{eq1}, which is here rewritten for the sake of clearness
\begin{equation}
H=H_{SSH}+H_{P}+H_{I},
\end{equation}
where
\begin{equation}
H_P= \frac{\omega_0}{2} \sigma_z
\end{equation}
is the Hamiltonian of the qubit with Rabi frequency $\omega_0$, 
\begin{eqnarray}
H_{SSH} & = & \omega_a \sum (a^{\dag}_n a_n+b^{\dag}_n b_n)    \\
& + & t_1 \sum_n (a^{\dag}_n b_n +hc)+ t_2 \sum_n(a^{\dag}_nb_{n-1}+hc) \nonumber
\end{eqnarray}
 is the SSH Hamiltonian, with the band gap center at energy $\omega_a$, and
 \begin{equation}
 H_I= \gamma[\sigma_+ (a_0+b_0)+ \sigma_{-} (a^{\dag}_0+b^{\dag}_0)]
 \end{equation}
 is the interaction term in the rotating-wave approximation. 
 In the above equations, $\sigma_z= |e \rangle \langle e |-|g \rangle \langle g |$, $\sigma_+=|e \rangle \langle g|$, $\sigma_-=|g \rangle \langle e|$ are the 
 Pauli pseudo spin operators associated to the qubit with ground and excited states $| g \rangle$ and $| e \rangle$, respectively, and $\gamma$ is the strength of the qubit-bath dissipative coupling. 
 Here we consider the case $\omega_0=\omega_a$,  corresponding to absence of decay in the weak-coupling (markovian) approximation, however we do not make any assumption on the coupling $\gamma$, which could be of the same order of magnitude or even larger than the hopping rates $t_{1,2}$ of the SSH chain, i.e. of the width of energy bands/gap (strong coupling regime). \\
 \\
  {\bf Bound states and decay dynamics.}
 Let us assume that at initial time $t=0$ the SSH chain is in the vacuum state, while the probing two-level system is prepared in the excited state, i.e. $| \psi(0)\rangle=| e \rangle \otimes |0 \rangle$. Apart from a phase term rotating at the frequency $\omega_0$, the time-evolved state can be written as 
 \begin{equation}
 | \psi(t) \rangle=  q(t) |e \rangle  \otimes |0 \rangle + | g \rangle \otimes  \sum_n \left( \alpha_n(t) a^{\dag}_n  + \beta_n(t) b^{\dag}_n  \right) | 0 \rangle
 \end{equation}

\begin{figure*}
\includegraphics[width=16cm]{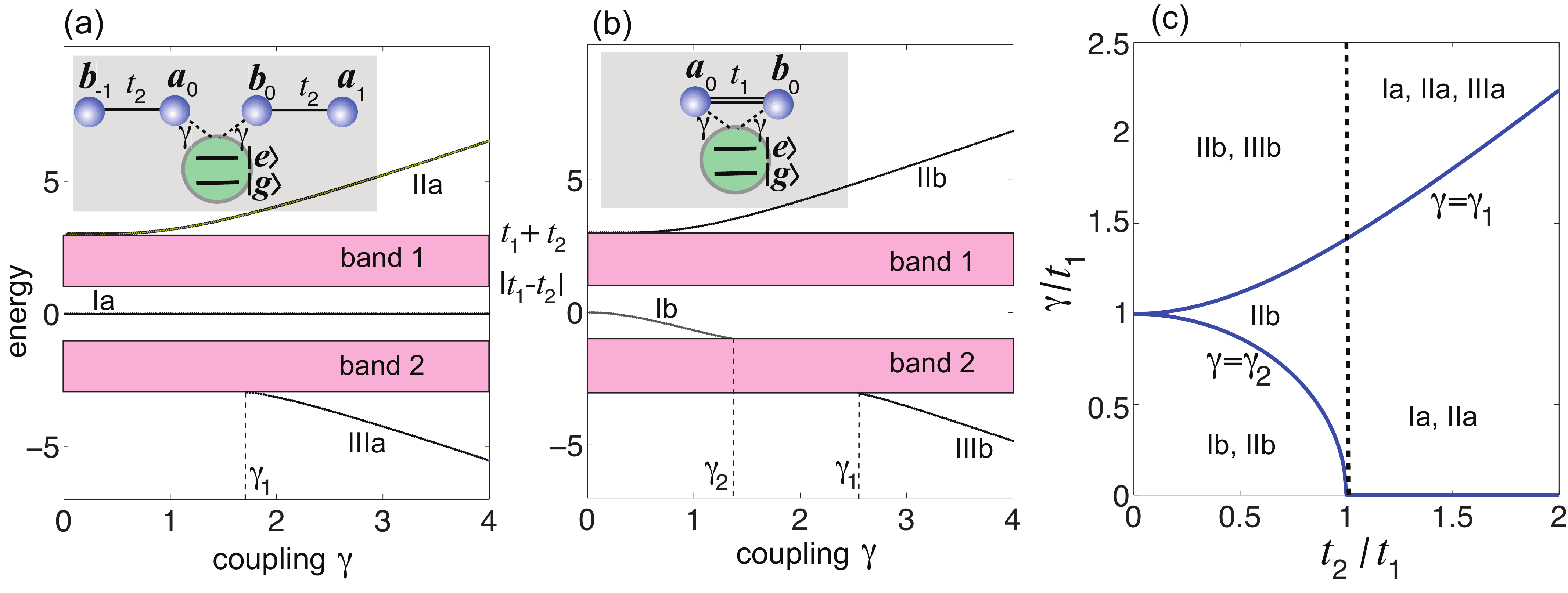}
\caption{(Color online)  (a,b) Energy spectrum (band diagram) of the Hamiltonian $\mathcal{H}$,  for (a) the topological phase $t_1<t_2$, and (b) $t_1>t_2$. In (a), there are three  bound states (BSs), denoted by Ia, IIa, and IIIa: a zero-energy topologically-protected state (Ia), and two BSs without topological protection that emanate from above and from below of the upper and lower bands. Mode IIa is thresholdless, whereas mode IIIa exists for $\gamma>\gamma_1 \equiv \sqrt{t_1(t_1+t_2)}$. In (b) all three modes do not have topological protection. Mode Ib is thresholdless but disappears for $\gamma>\gamma_2 \equiv  \sqrt{t_1(t_1-t_2)}$, mode IIb is thresholdless and does exist for any value of coupling $\gamma$, whereas mode IIIb exists for $\gamma>\gamma_1\equiv \sqrt{t_1(t_1+t_2)}$. Parameter values used in the plots are $t_1=1$, $t_2=2$ in (a), and $t_1=2$, $t_2=1$ in (b). The insets show the coupling of the qubit with the SSH chain in the flat band limit [$t_1=0$ in (a) and $t_2=0$ in (b)], where only few sites of the lattice can be populated and the number of bound states [three in (a) and two in (b)] can be readily derived from symmetry considerations. (c) Domain of existence of BSs  of the Hamiltonian $\mathcal{H}$ in the $(t_2/t_1, \gamma / t_1)$ plane. The vertical dashed line separates the two topological phases $t_1<t_2$ and $t_1>t_2$ of the SSH lattice.}
\label{palle1}
\end{figure*} 
 where the amplitudes $q(t)$, $\alpha_n(t)$ and $\beta_n(t)$ satisfy the coupled evolution equations
 \begin{eqnarray}
 i \frac{d \alpha_n}{dt} & = & t_2 \beta_n+ t_1 \beta_{n-1} +\delta_{n,0} \gamma q \label{alfa}\\
 i \frac{d \beta_n}{dt}& = & t_2 \alpha_n+t_1 \alpha_{n+1}+\delta_{n,0} \gamma q \label{beta}\\
 i \frac{dq}{dt} & = & \gamma(\alpha_0+\beta_0)\label{q}
 \end{eqnarray}
 with the initial conditions $\alpha_n(0)=\beta_n(0)=0$, $q(0)=1$.
Equations (\ref{alfa}--\ref{q})  can be written in compact form as
 \begin{equation}
 i \frac{d}{dt} \left( 
 \begin{array}{c}
 \alpha_n \\
 \beta_n \\
  q
  \end{array}
  \right)= \mathcal{H}
  \left(
  \begin{array}{c}
 \alpha_n \\
 \beta_n \\
  q
  \end{array}
  \right)
 \end{equation}
  where $\mathcal{H}$ is the Hamiltonian of the system in the single-excitation sector. Clearly, $|q(t)|^2$ is the probability that the two-level system remains in the excited state at time $t$. More generally, if at initial time the qubit is in a mixed state described by the density matrix $\rho_{i,k}=\langle i | \rho(0) | k \rangle$ ($i,k=g,e$) and the SSH lattice is in the vacuum state, the reduced density matrix $\rho(t)$  of the qubit at time $t$, obtained after tracing out the lattice degrees of freedom, reads
  \begin{equation}\label{rhodiss}
  \rho(t)= \left(
  \begin{array}{cc}
  |q(t)|^2 \rho_{ee} & q(t) \rho_{eg} \\
  q^*(t) \rho_{ge} & (1-|q(t)|^2) \rho_{ee}+\rho_{gg}
  \end{array}
  \right)
  \end{equation}
 Therefore, the coherence of the qubit decays as $\sim |q(t)|$, while the population as $\sim |q(t)|^2$. The temporal decay law of $q(t)$ can be formally written as contour integral in complex plane after solving Equations (\ref{alfa}--\ref{q}) using Laplace transform (or Green-function) method.  After some lengthy but straightforward calculations, one obtains
 \begin{equation}
 q(t)=\frac{1}{2 \pi i} \int_{- \infty+i 0^+}^{\infty+i0^+}  d \omega \frac{\exp(-i \omega t)}{-\omega+\frac{\gamma}{\pi} \int_{-\pi}^{\pi} dk \left[  \frac {\cos^2 \varphi(k)} {\omega-E_+(k)}+\frac{\sin^2 \varphi(k)}{\omega-E_-(k)} \right] }
 \end{equation} 
where $E_{\pm}(k)= \pm H(k)= \pm \sqrt{t_1^2+t_2^2+2 t_1 t_2 \cos k}$ are the energy dispersion curves of the two SSH bands, while the phase $\varphi(k)$ and amplitude $H(k)$ are defined by the 
 relation $H(k) \exp[i \varphi(k)/2] \equiv t_2+t_1 \exp(-ik)$.
 After an initial transient, the behavior of $q(t)$ is determined by the poles of the Green function, which correspond to BSs of the Hamiltonian $\mathcal{H}$. Therefore, the decay behavior of the coherence $\rho_{eg}(t)=q(t) \rho_{eg}$ is ultimately determined by the number and frequencies of the BSs of the Hamiltonian $\mathcal{H}$.\\
 Figures \ref{palle1}(a) and (b) show the energy spectrum of the Hamiltonian $\mathcal{H}$ versus the coupling $\gamma$  for the two topological phases $t_1<t_2$ and $t_1>t_2$ of the SSH lattice. The energy spectrum clearly comprises a continuous spectrum which reproduces the two bands of the SSH lattice, separated by the gap $|t_1-t_2|$ and independent of the topological phase of the lattice, and a number of BSs, localized near $n=0$, which depend on the topological phase of the lattice. Clearly, for symmetry reasons the BSs can be classified as even (symmetric) and odd (anti-symmetric) modes, corresponding to $\beta_{-n}=\pm \alpha_{n}$. It can be readily shown that BSs with odd symmetry ($\beta_{-n}=-\alpha_{n}$) do not exist, while even-symmetric BSs ($\beta_{-n}=\alpha_{n}$) can be searched by making the Ansatz
 \begin{eqnarray}
 \alpha_n  = c_1 X^{-n} \exp(-iEt) & \; \; n \geq 1 \\
 \beta_n =  c_2 X^{-n} \exp(-iEt)  & \; \; n \geq 0 \\
 q =  q_0 \exp(-i Et) 
 \end{eqnarray} 
 In the above equations, $E$ is the eigenenergy of the BS, $X$ is related to its localization length (with $|X|>1$ for localization), $c_{1,2}$ are constant amplitudes that determine the relative occupation of the the excitation in the two sublattices of the SSH chain, and $q_0$ is the occupation amplitude of the qubit. The above Ansatz provides an eigenstate of the Hamiltonian $\mathcal{H}$ provided that the following conditions are met 
  \begin{eqnarray}
 E c_1 & = &  (t_1+t_2X) c_2 \label{ec1}\\
 E c_2 & = & (t_1+t_2/X) c_1 \label{ec2}\\
 (E-t_1) c_2 & = & c_1 t_2/X+ \gamma q_0 \\
 E q_0 & = & 2 \gamma c_2.
 \end{eqnarray}
 From Eqs.(\ref{ec1}) and (\ref{ec2}), it readily turns out that the energy $E$ of the BS is related to the localization parameter $X$ by the dispersion relation
 \begin{equation}
 E^2=t_1^2+t_2^2+t_1t_2 (X+1/X)\label{e2}
 \end{equation}
while the amplitudes $c_2$ and $q_0$ are the solutions to the homogeneous system of equations
\begin{eqnarray}
(t_1^2+t_1t_2X-Et_1) c_2- \gamma E q_0 =0 \\
2 \gamma c_2-E q_0=0.
\end{eqnarray}
The solvability condition of the homogeneous system yields either $E=0$ or
\begin{equation}
E=t_1+t_2X-2 \gamma^2/t_1.\label{eqe}
\end{equation}
For  $E=0$ one obtains $c_2=0$, $X=-t_2/t_1$, $q_0/c_1=t_1 / \gamma$. The localization condition $|X|>1$ indicates that the zero-energy BS exists in the topological phase $t_2>t_1$ solely. Note that this state shows topological protection, i.e. its energy does not change as the ratio $t_1/t_2$ is varied until the gap is closed.
The other possible values of energies are obtained by solving the nonlinear system of equations (\ref{e2}) and (\ref{eqe}). After elimination of the energy $E$, the following cubic equation for the localization parameter $X$ is obtained
\begin{equation}\label{eqx3}
X^3+s_1X^2+s_2X+s_3=0
\end{equation}
with coefficients
\begin{eqnarray}
s_1 & = & \frac{t_1}{t_2}-4 \frac{\gamma^2}{t_1 t_2} \\
s_2 & = & -1-4 \frac{\gamma^2}{t_2^2}+4 \frac{\gamma^4}{t_1^2t_2^2} \\
s_3 & = & -\frac{t_1}{t_2}.
\end{eqnarray}
 \begin{figure}
\includegraphics[width=8.4cm]{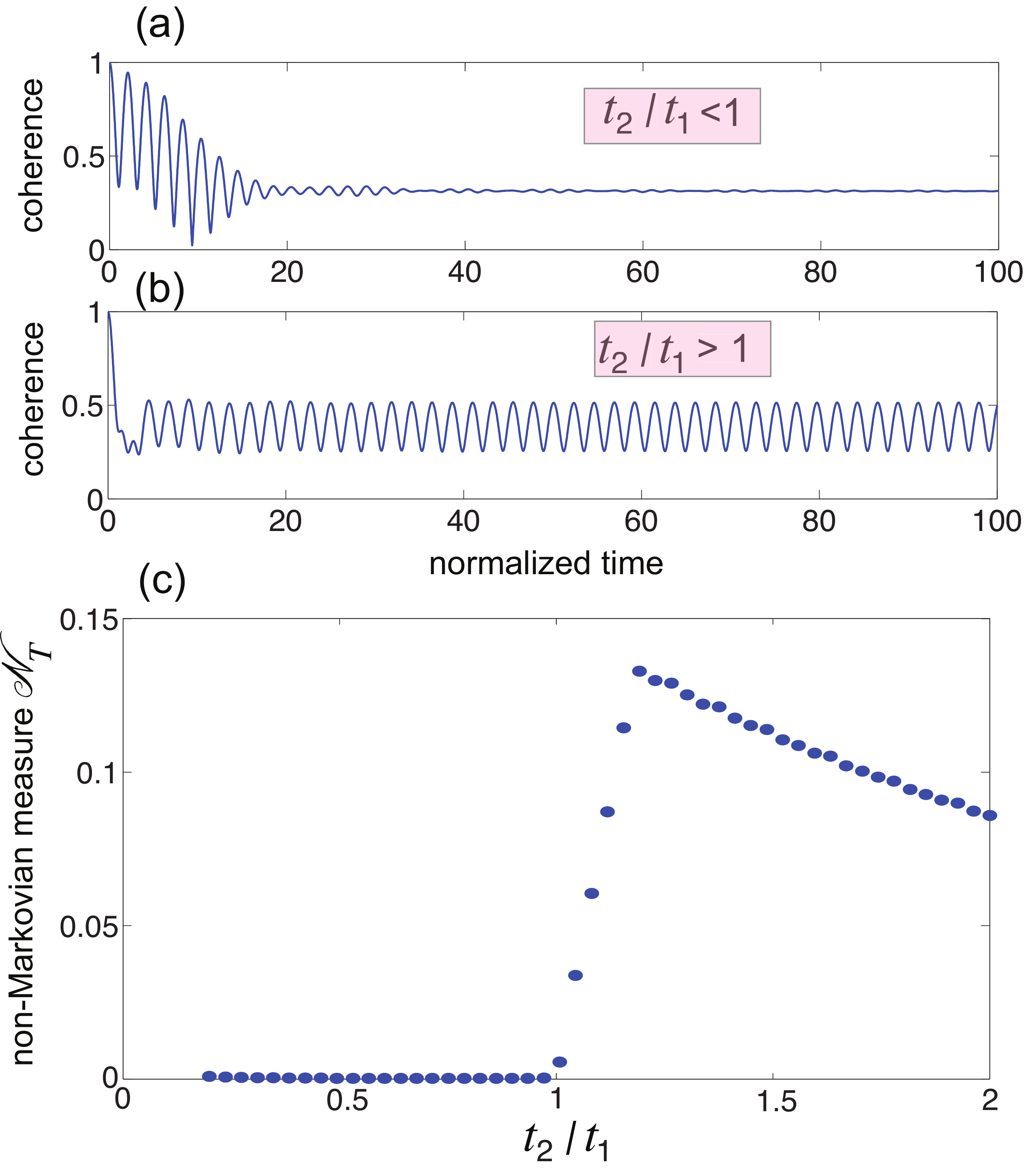}
\caption{(Color online) Decay behavior of the coherence $|q(t)|$ for (a) $t_2/t_1=0.3$, and (b) $t_2/t_1=1.5$. (c) Behavior of the non-Markovianity measure $\mathcal{N}_T$ versus $t_2/t_1$. In the simulations we assumed $\gamma=t_1$ and time is in units of $1/t_1$.}
\label{palle2}
\end{figure} 
The acceptable roots of Equation (\ref{eqx3}) are those with $|X|>1$ (for localization of the BS) and $X$ real (because the energy $E$ should be real). Once an acceptable root $X$ of the cubic equation has been found, the corresponding energy eigenvalue is obtained from Equation(\ref{eqe}).\\ 
The boundaries for existence of BSs in the $(t_2/t_1, \gamma / t_1)$ can be readily found by  setting $X= \pm 1$ in Equation (\ref{eqx3}). The states with $X=1$ correspond to emergent BSs that emanate from the top (or the bottom) of the upper (or lower) band, whereas  the states with $X=-1$ emanates from the bottom (or the top) of the upper (or lower) band. The domain of existence of BSs and typical band diagrams for the two different topological phase $t_1<t_2$ and $t_2>t_1$ are illustrated in Figure\ref{palle1}. As shown above, in the topological phase $t_1<t_2$  there is always one BS at zero energy (denoted by Ia in the figure), which shows topological protection likewise an edge state of the SSH chain. There are also other two edge states without topological protection, one being thresholdless (mode IIa) while the other one (mode IIIa) emerges from the bottom of the lower band at $\gamma> \gamma_1 \equiv \sqrt{t_1(t_1+t_2)}$; see Figure\ref{palle1}(a). In the other topological phase $t_1>t_2$ one can find upmost two BSs without topological protection. A first BS (mode Ib) is thresholdless and falls into the continuum of lower band at $\gamma>\gamma_2 \equiv \sqrt{t_1(t_1-t_2)}$. The second BS (mode IIb) is tresholdless and its energy lies above the upper band, while the third BS (mode IIIb) emerges from the bottom of the lower band at $\gamma>\gamma_1$; see Figure\ref{palle1}(b). Note that in the limit of flat bands $t_1 \rightarrow 0$ or $t_2 \rightarrow 0$ in the two topological phases, there are three and two BSs, respectively. The different number of BSs stem from the fact that, in the flat band limit, the qubit probes a finite but different number of sites of the SSH chain, as schematically shown in the insets of Figure\ref{palle1}(a) and (b). In Figure\ref{palle1}(a), there are five sites overall that can be occupied, corresponding to five states, two with odd symmetry and the other three with even symmetry. Since only the even modes have excitation in the qubit, one concludes that there are three BSs relevant for the qubit decay dynamics. On the other hand, in Figure\ref{palle1}(b) there are only three sites overall that can be occupied, corresponding to three states with defined symmetry. Only the two states with even symmetry have excitation in the quibit, and therefore in this case we have two BSs relevant for the decay dynamics of the qubit.

\begin{figure}
\includegraphics[width=8.3cm]{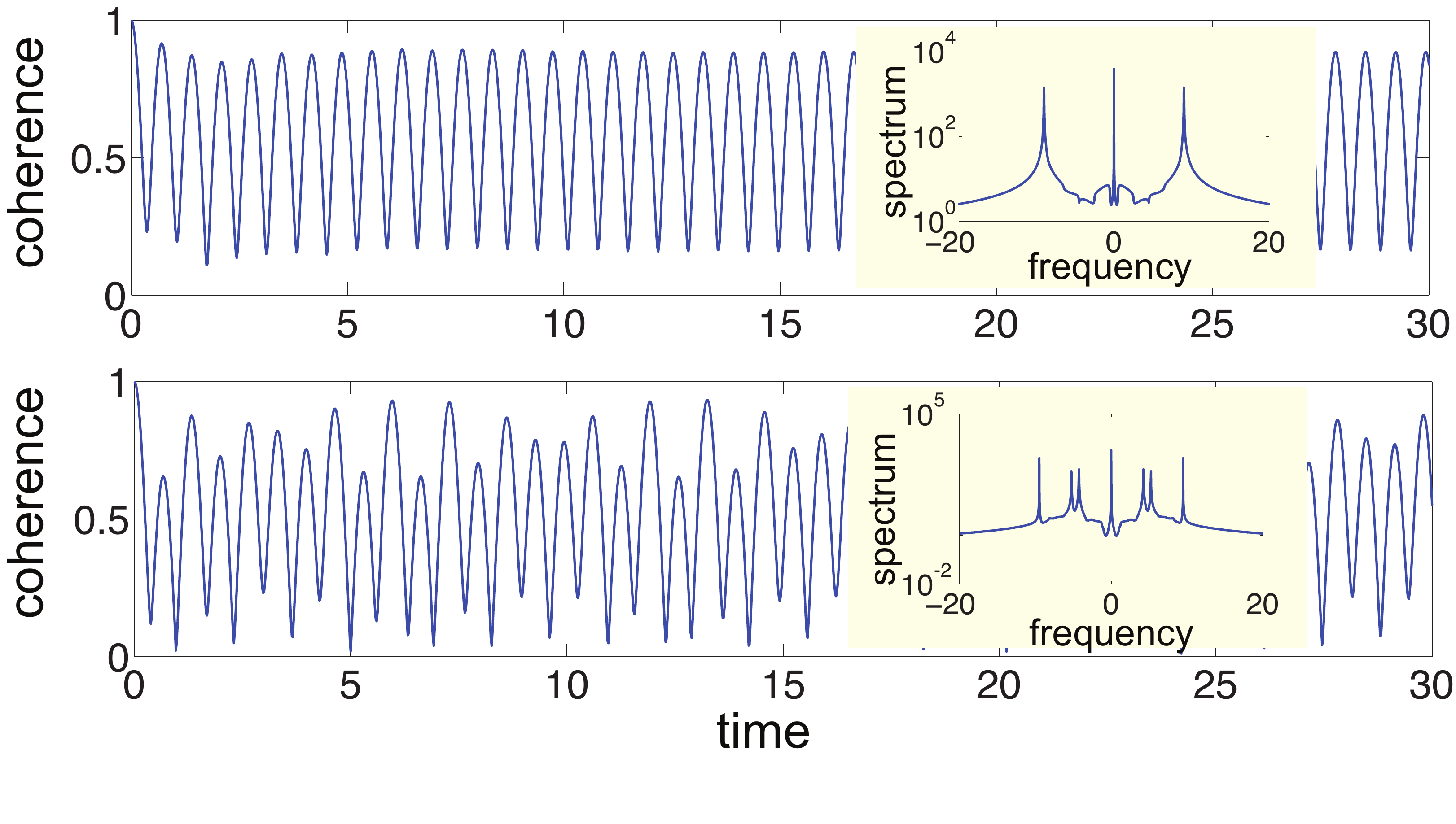}
\caption{(Color online) Numerically-computed behavior of the coherence $|q(t)|= \sqrt{L(t})$ for (a) $t_1=2$, $t_2=1$, $\gamma=3$, and (b) $t_1=1$, $t_2=2$, $\gamma=3$. The insets show the spectrum (in arbitrary units and on a log scale) of  $L(t)$. Note the different numbers of peaks in the frequency spectrum in (a) and (b), related to the different numbers of BSs in the two topological phases $t_1>t_2$ and $t_1<t_2$.}
\label{palle3}
\end{figure} 
{\bf Non-monotonicity of the coherence decay and non-Markovianity measure.}
The decay dynamics of the coherence $|q|$  comprises an initial transient decay followed by a non-decay dynamics which can be oscillatory or non-oscillatory depending on the number of BSs of the Hamiltonian $\mathcal{H}$, as in the dephasing coupling scheme discussed in the main text.   The most clear signature of topological phases on non-Markovianity is obtained when the coupling $\gamma$ of the qubit with the SSH bath is tuned to equal $t_1$, i.e. for $\gamma=t_1$ . As $t_2$ is varied from below to above $t_1$, so as the gap closes and the SSH lattice undergoes a topological phase transition, the decay of coherence shows a qualitative change, as depicted in Figure\ref{palle2}. In the topological phase $t_1<t_2$, there are two BSs so as the asymptotic behavior of {$|q(t)|$}  is  oscillatory at a frequency given by the energy separation of the two BSs [see Figure\ref{palle2}(b)]. On the other hand, in the topological phase $t_1>t_2$ there is only one BS and, after an initial transient, the {coherence} settles down to a steady-state (non-oscillatory) value [Figure\ref{palle2}(a)]. 
As discussed in the main text  and according to  \cite{nm},  non-Markovianity  can be quantified through the lack of contractiveness of the dynamical map. Given the map  of \Eq{rhodiss}, ${\cal N}$ is fully determined by the 
increase of the coherence (recoherence) time windows of $|q(t)|$ \cite{xu}. As discussed in the pure dephasing scenario, the oscillatory behavior of $\rho(t)$ would lead to a divergence of non-Markoviantity, whose behavior can be regularized introducing the temporal average
 \begin{equation}
{\mathcal N}_T= \frac{1}{T} \int_{0 \; \dot{|q|}>0}^{T} dt \frac{d 
|q(t)|}{dt},
\end{equation}
where the integral is extended over the time intervals where the derivative  { $(d|q|/dt)$} is positive. For long times $T$ it turns out that $\mathcal{N}_T$ vanishes in the topological phase $t_1>t_2$, while it reaches a steady and non-vanishing value in the $t_1<t_2$ topological phase.  The vanishing of non-Markovianity for a unique value of the system parameters was also reported in a different context in ref. \cite{apollaro}.
 Therefore, the non-Markovianity measure $\mathcal{N}_T$ provides an order parameter of the topological quantum phase transition.
 This is shown in Figure\ref{palle2}(c), where we numerically computed the non-Markovianity measure $\mathcal{N}_T$,  after an initial transient decay dynamics, versus the ratio $t_2/t_1$ with $\gamma=t_1$.\\ 
For a coupling $\gamma$ different than $t_1$, $\mathcal{N}_T$ does not provide anymore an order parameter of the topological quantum phase transition. However, provided that the coupling $\gamma$ is strong enough, namely for $\gamma>\gamma_1$ (strong coupling regime), different topological phases of the SSH bath result in a qualitatively different behavior of the coherence, as shown in Figure\ref{palle3}.
 In the topological phase $t_1>t_2$, there are two BSs and therefore the coherence $|q(t)|$ shows a single-frequency oscillatory dynamics [Figure\ref{palle3}(a)], which is clearly visible in the three-peaks structure of the frequency spectrum of $L(t)$ [see the inset of Figure\ref{palle3}(a)].  On the other hand, in the topological phase $t_1<t_2$, there are three BSs and therefore the coherence $|q(t)|$ shows a multi-frequency and rather generally aperiodic oscillatory dynamics [Figure\ref{palle3}(b)], which is clearly visible in the seven-peaks structure of the frequency spectrum of $L(t)$ [see the inset of Figure\ref{palle3}(b)].  To conclude, in the dissipative coupling scheme the non-Markovianity measure $\mathcal{N}_T$ provides a topological order parameter only for $\gamma=t_1$. When such a condition is not satisfied, still different topological phases of the SSH bath can be distinguished looking at the Fourier spectrum of the coherence, provided that we operate in the strong coupling regime ($\gamma>\gamma_1$).

\section{ Dephasing coupling: bound states}\label{appB}
\renewcommand{\thesubsection}{B}
\renewcommand{\theequation}{B-\arabic{equation}}
\setcounter{equation}{0}
\renewcommand\thefigure{B-\arabic{figure}} 
\setcounter{figure}{0}


In this section, we will mainly focus on the analytical calculation of the BSs for the dephasing model of Eqs. (1-3). Starting from the states defined by Equation (6)  in the main text, we look for possible solutions to the Schr\"odinger equations $\bar H\ket{\psi_\pm}=E_\pm\ket{\psi_\pm}$, with $\bar H=H_{SSH}+ \gamma (a_{0}^\dag a_{0}+b_{0}^\dag b_{0})$.  One readily obtains
\begin{eqnarray}\label{sch}
\bar H\ket{\psi}_\pm &=& \{\tilde\gamma_\pm  c_0 (a_{0}^\dag\pm b_{0}^\dag ) + c_0 t_2(b_{-1}^\dag\pm a_{1}^\dag )+\nonumber\\
&&+ c_1 [t_2(a_{0}^\dag \pm b_{0}^\dag )+t_1 (a_{-1}^\dag \pm b_{1}^\dag ) ]+\dots\}\ket{0}=E_\pm\ket{\psi}_\pm,\nonumber\\
\end{eqnarray}
where $\tilde\gamma_\pm=\gamma\pm t_1$.
Notice that solving  these equations is equivalent to taking as an effective Hamiltonian a semi-infinite chain obtained by cutting the original one in the middle of the impurity cell and replacing the impurity energy with $\tilde\gamma_\pm$. This renormalization is due to a ``bounce" around the impurity. In the limit of a chain much  longer than the localization of the BSs, the dimerically tailored Ansatz $ c_{0}=X^n c_{2 n} $ and  $ c_{2n+1} = Y c_{2 n}$ allows one to write two sets of closed equations
\begin{equation}\label{ep1}
\begin{cases} 
\tilde\gamma_\pm t_1+Y\;t_2=E_\pm\\
 t_2+ t_1/X=E_\pm\;Y\\
t_2+ t_1\;X= E_\pm /Y. \end{cases}
\end{equation}
The quantity $X$ determines the localization length of the BSs expressed in cell units, and
the condition  $|X|>1$  should be satisfied for localization. In other words, solutions of (\ref{ep1}) corresponding to $|X|< 1$ are not  BSs of (\ref{sch}) and must be discarded. The explicit solutions (two for any of the systems) for the energies 
are the following:
\begin{equation}\label{ep}
E_+^{\pm}=\frac{\gamma ^2+2 t_1^2+2 \gamma  t_1\pm\sqrt{\left(\gamma ^2+2 \gamma  t_1\right)^2+4 t_2^2 (\gamma +t_1)^2}}{2 (\gamma +t_1)}
\end{equation}
 and
 \begin{equation}\label{em}
E^\pm_-=\frac{\gamma ^2+2 t_1^2-2 \gamma  t_1\pm\sqrt{\left(\gamma ^2-2 \gamma  t_1\right)^2+4 t_2^2 (\gamma - t_1)^2}}{2 (\gamma -t_1)}
\end{equation}
 while the respective localization length  parameters are
 \begin{eqnarray}
 X_+^{\pm}&=&\frac{\gamma ^2+2 \gamma  t_1\pm\sqrt{\left(\gamma ^2+2 \gamma  t_1\right)^2+4 t_2^2 (\gamma +t_1)^2}}{2 t_1 t_2}\nonumber\\
 X_-^\pm &=&\frac{\gamma ^2-2 \gamma  t_1\pm\sqrt{\left(\gamma ^2-2 \gamma  t_1\right)^2+4 t_2^2 (\gamma - t_1)^2}}{2 t_1 t_2}\nonumber\\
 \end{eqnarray}
 and the dimer imbalances are 
 \begin{eqnarray}
  Y_+^{\pm}&=&\frac{\gamma+t_1}{t_1}\frac{1}{X_+^{\pm}}\nonumber\\
   Y_-^{\pm}&=&\frac{\gamma-t_1}{t_1}\frac{1}{X_-^{\pm}}.
 \end{eqnarray}
 Finally, the normalization condition of states $\ket \psi_\pm$ determines the value of $c_0$, which reads
\begin{equation}
c_0=\sqrt{\frac{1-(1/X)^{2}}{2(1+Y^{2})}}.
\end{equation}
 for any of the four  solutions.
 In \Fig{figx}, we plot the inverse of the localization length for a fixed value of $\gamma$ as a function of $t_2/t_1$, that, together with \Fig{figbands} of the main text, gives us the complete picture about the existence of BSs detailed in the main text. 
 
 \begin{figure}
 \begin{center}
 \includegraphics[scale=.55]{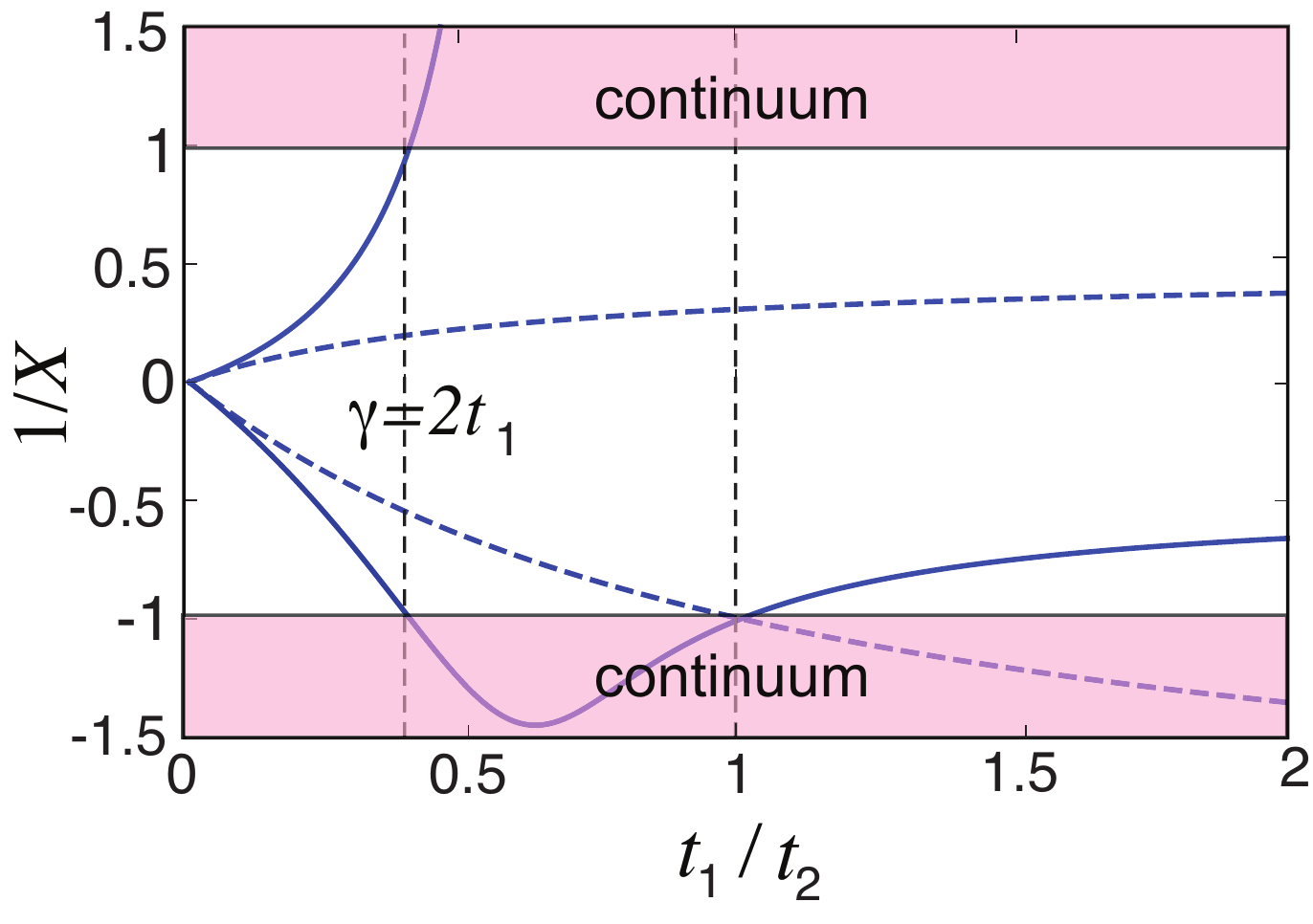}
\caption{ Inverse of the localization length parameter and conditions for the existence of the BSs. 
The curves display symmetric solutions $1/X_+$  (dashed) and antysimmetric ones $1/X_-$ (solid) as a function of $t_1/t_2$ for $\gamma=0.8 t_2$. 
Three phases can be clearly defined: for $t_1/t_2<\gamma/2$, all the quantities are smaller than one and four BSs are found; 
for $\gamma/2<t_1/t_2<1$ there are only two symmetric BSs, while for $t_1<t_2$ there are one symmetric and one anti-symmetric BS.   }
\label{figx}
 \end{center}
\end{figure}

\begin{figure}[t]
\includegraphics[scale=.32]{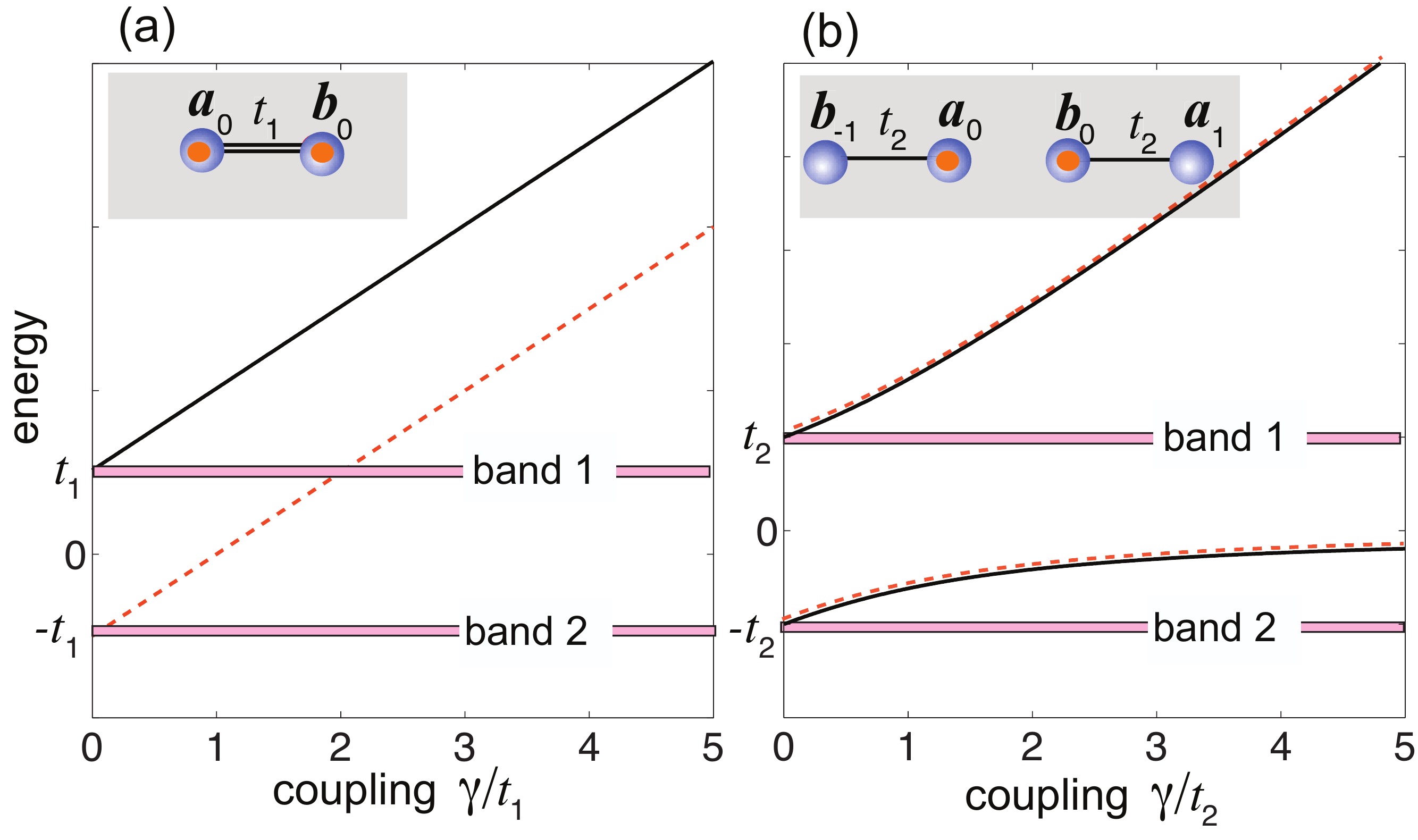}
\caption{ Energy spectrum of $\bar{H}$ as a function of the coupling $\gamma$ in the flat band limit (a) $t_2 \rightarrow 0$, and (b) $t_1 \rightarrow 0$. The bands shrink to two flat lines, while only few sites of the lattice [two in (a) and four in (b)] interact with the qubit. In (a) there are two BSs, one symmetric (solid line) and the other one anti-symmetric (dashed line), while in (b) there are four BS, two symmetric and the other two anti-symmetric, which are degenerate in energy.}
 \label{basta}
\end{figure}

The reason why in one case there are always two BSs and in the other one there can be  four of them can be understood looking at the flat-band limit (Figure\ref{basta}). In the phase $t_1>t_2$, this limit corresponds to taking $t_2=0$, which means that the impurity cell is completely decoupled from the rest of the dimers and can be described by the cell Hamiltonian $H_{0}^{t_2=0}=t_1(a_{0}^\dag b_{0}+b_{0}^\dag a_{0})+\gamma (a_{0}^\dag a_{0}+b_{0}^\dag b_{0})$, which admits the pair of eigenstates  of opposite parity $(a_{0}^\dag\pm b_{0}^\dag)\ket{0}$ with eigenvalues $\gamma\pm t_1$ [Figure\ref{basta}(a)]. On the other hand, in the other phase we would have $t_1=0$. Then, the Hamiltonian around the impurity would be $H_{0}^{t_1=0}=t_2(a_{0}^\dag b_{-1}+b_{0}^\dag a_{1}+h.c.)+\gamma (a_{0}^\dag a_{0}+b_{0}^\dag b_{0})$, which correspond to two disconnected dimers, each of them admitting two eigenstates of different parity [Figure\ref{basta}(b)]. The transition from four to two BSs cannot be analyzed using the flat-band argument, as it only happens for finite values of $t_1$.

\end{document}